\documentclass[ifip]{svmult}
\usepackage{setspace}
\usepackage{longtable}
\usepackage{graphicx}
\usepackage{url}
\newcommand{\tra}{{\tt Trac}}
\newcommand{\dpr}{{\tt DrProject}}
\date{}

\begin{document}

\title{Code forking in open-source software: a requirements perspective}

\author{Neil A. Ernst, Steve Easterbrook and John Mylopoulos}
\institute{
 Department of Computer Science\\University of Toronto, Toronto, Canada\\
{\tt \{nernst,sme,jm\}@cs.utoronto.ca}}

\maketitle
\thispagestyle{empty}

%********************************************************************
%* Copyright notice
%********************************************************************
% \renewcommand{\thefootnote}{}
% \footnotetext{\footnotesize
% Copyright \copyright\ 2006 by the authors. 
% Permission to copy is hereby granted provided the original copyright
% notice is reproduced in copies made.}
% \renewcommand{\thefootnote}{\arabic{footnote}}

\begin{abstract}
To fork a project is to copy the existing code base and move in a direction different than that of the erstwhile project leadership. Forking provides a rapid way to address new requirements by adapting an existing solution. However, it can also create a plethora of similar tools, and fragment the developer community. Hence, it is not always
clear whether forking is the right strategy. In this paper, we describe a mixed-methods exploratory case study that investigated the process of forking a project. The study concerned the forking of an open-source tool for managing software projects, {\tt Trac}. \tra\ was forked to address differing requirements in an academic setting. The paper makes two contributions to our understanding of code forking. First, our exploratory study generated several theories about code forking in open source projects, for further research. Second, we investigated one of these theories in depth, via a quantitative study. We conjectured that the features of the OSS forking process would allow new requirements to be addressed. We show that the forking process in this case was successful at fulfilling the new projects requirements.
\end{abstract} 

\section{Introduction}
\label{intro}
Traditionally, requirements analysis constitutes the initial phase of a software development project, and implementation the last. But what happens when a codebase already exists and serves as the starting point of software development? We study this question in the context of open source software. Open-source software (OSS) has proved its worth. OSS runs some of the largest web sites in the world (Apache), complex corporate servers (Linux/BSD), and is comparable to proprietary desktop application suites (OpenOffice) \cite{weber04}.  However, there remains a perception that OSS is not high-quality, and adoption of OSS in corporate IT environments is still low. Goode \cite{goode05}, in his survey of Australian corporations, discusses several reasons for this.  They include a perceived lack of support, hidden installation costs, and finally, a belief that corporate IT requirements are not met by any existing OSS tools.  
%However, there are OSS tools that partially fulfill these requirements; furthermore, the advantage of an open licence is that anyone may derive new products from that software. 

To fork a project is to copy the existing code base and move in a direction different than that of the erstwhile project leadership. Forking the codebase allows developers to leverage existing functionality while also addressing new requirements. Although flexible, forking has inherent difficulties, such as maintenance, evolution, and social factors concerning the development community. This paper looks at the form these take. To study the problem and evaluate potential solutions we focus on a software project management tool, {\tt Trac}. Inasmuch as {\tt Trac} had an original target audience, it was intended for developers in a corporate environment, then released more widely. In 2005 the tool was forked to support the development of a project that targeted undergraduate programming projects. This process, and its implications, are the subject of this research.

We followed a two-phase, mixed-methods sequential approach \cite{creswell02} examining the interaction of forking and requirements. In the exploratory phase (Section \ref{qual}), we qualitatively examined the nature of OSS, requirements, and forking by conducting surveys and interviews of participants from both the original and the forked project. Using grounded theory \cite{strauss98} methodology, we generated several theories about the problem. 

We then selected one theory and applied confirmatory analysis in the second phase. In this primarily quantitative segment, the theory we tested was whether some of the different requirements that motivated the fork were met. We analyzed this using two non-functional requirements. We first used a variety of code metrics of the two codebases (Section \ref{quant}) to assess {\em maintainability}. Secondly, we used qualitative usability analyses to assess interface complexity, or {\em usability}.

We designed this study as an exploratory one since there was a lack of detailed research regarding the forking of OSS to meet differing requirements. This way we generate some suitable research hypotheses for further testing. The second, more quantitative portion of the research examines whether code forking can successfully address certain new requirements. In some respects, we argue in Section \ref{implications}, many of the theories we initially generated are best suited to qualitative or case study analysis.

Our contribution is two-fold. One, we describe some of the process issues that arise when forking OSS. We provide theory patterns derived from the data that are a step towards more comprehensive theories about this process. Two, we describe a case study on a product of a fork, {\tt DrProject}. In this case study we examined how well the forked tool could be adapted to meet new requirements, that didn't exist in the original.

\section{Background}
\label{back}
We begin with a brief introduction to the tools involved.  To provide context, we then briefly review the literature. This research adds to the existing literature because few people have looked at open-source software from a requirements perspective (requirements are rarely mentioned in OSS; often a project is initiated due to perceived need). As well, there has been little work on the process of forking.

\subsection{Context: {\tt DrProject} and {\tt Trac}}
\label{argon}
The context of this study is the evolution of a software project management tool named {\tt Trac}\footnote{\url{http://projects.edgewall.com/trac}}.  {\tt Trac} is currently (as of January 2006) in version 0.9.3 and supports, among other things, Subversion repository integration, Wiki-enabled web pages, tickets and bug filing (a ticket is a to-do item assignable to project members).  {\tt Trac} is developed as an open-source project, under the leadership of two developers at Edgewall Software.  

In January of 2005 two faculty members at the University of Toronto (not involved in conducting this research) decided to fork {\tt Trac} in order to support undergraduate programming. {\tt Trac} was chosen, according to interviews with the proponents, for three reasons: 1) it was written in a familiar language (Python); 2) it had simple technology (CGI); 3) it had a small codebase and clean UI. The result of the fork, {\tt DrProject}\footnote{\url{http://pyre.third-bit.com/drproject}}, is the latest in a series of attempts to develop software to support student programming projects at the undergraduate level. Educators at the University of Toronto believed graduating students were leaving unaware of the importance of bug reports and version control. They determined that a tool should be developed to expand that pedagogical objective (notwithstanding reports that a large number of corporate software developers were equally unfamiliar with the tools). 

\begin{figure*}[htb]
\begin{center}
\includegraphics[bb=0 0 546 216,width=3in]{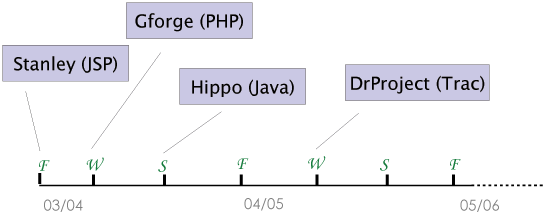}
\caption{A short history of the incarnations of {\tt DrProject}. Letters indicate school terms (Fall, Winter, Summer)}
\label{fig:argon-timeline}
\end{center}
\end{figure*}

\begin{figure*}[htb]
\begin{center}
\includegraphics[bb=0 0 716 650,height=3in]{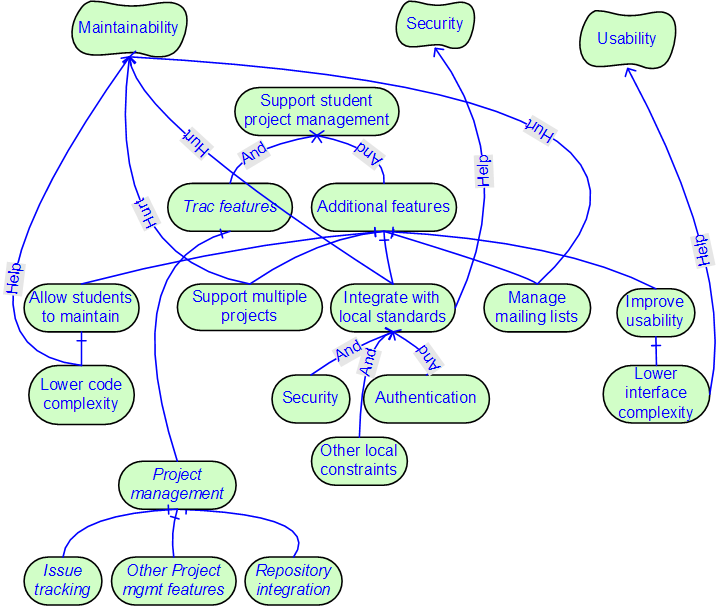}
\caption{Goal decomposition for {\tt DrProject}. Notation used is based on the NFR framework \cite{mylopoulos92}. Goals addressed by the existing {\tt Trac} tool are in italics. Other goals were extracted from qualitative studies conducted as part of this research.}
\label{fig:goal}
\end{center}
\end{figure*}

There have been a a number of previous attempts to develop such a tool at the University of Toronto (refer to Fig. \ref{fig:argon-timeline}). This latest iteration has been the most successful, with adoption by two separate courses, a self-hosted development project, and interest from academic research labs in the Department. We elicited the key requirements through interviews with the proponents. Functional requirements include, in addition to what \tra\ offers, multiple project support, test suite integration, and mailing list management. Non-functional requirements are for a codebase that is maintainable by students, and an interface that contains just those features that are necessary for short-duration software projects. These `early' requirements are presented in the goal graph of Fig. \ref{fig:goal}. 

\subsection{OSS, requirements and forking}
Research into the nature of open-source software has spawned a large degree of interest from a variety of disciplines, including economics, management, sociology, and software engineering~\cite{weber04,feller00}. Source code accessibility opened the door to rich stores of data on software evolution, development styles, and community building, {\em e.g.}, \cite{german04,godfrey00}. There seems to be scant empirical work on the nature of forking in OSS, beyond discussions of psychological and sociological issues at play (for example, motivations for contributing to an OSS project). Most literature mentions forking as a pitfall, or something to be avoided {\em e.g.} \cite{fitzgerald04}. There is mention of the process in several books, but it is mostly conjecture or anecdotal in nature, {\em e.g.}, \cite{raymond01}. One useful distinction, albeit conjectured, is presented by Fogel \cite{fogel05}. He distinguishes between forks that happen due to fundamental differences in project goals (as is the case here), and those which are custody battles over the existing project.

In this paper, we use the term to refer to situations where code from one project is duplicated in a different repository, and significant new requirements are added to the new project. This requirements-centric definition excludes downloaded code that receives minor alterations (e.g., a research tool for gene expression data). We are most interested in code that is altered to address new requirements. That is clearly the case in this project. 

There has been some research on requirements in OSS. Scacchi \cite{scacchi02} looks at how requirements are handled in different OSS projects and provides a good look at differences between OSS communities and other models. For example, he makes the observation that in OSS, requirements are tightly integrated with tool design and implementation. Scacchi {\em et al.} \cite{scacchi04} expand on the earlier paper with an ethnographic analysis of the requirements process in the NetBeans product. Our work differs in its focus on adapting existing tools to new requirements, rather than creating new tools based on (possibly emergent) requirements. Regarding the question of adapting OSS for corporate requirements, Bonaccorsi and Rossi \cite{bonaccorsi05} present a survey that suggests most corporate users of OSS adapt the tool to meet customer requirements without returning their contributions to the original project. Adams {\em et al.} \cite{adams05} take a similar tack, but they describe how they adapt a tool using a plugin architecture, rather than working at the internal level of design, as was the case here. Finally, similar themes appear in fairly extensive research into COTS integration with project requirements, for example, \cite{maiden98c,morisio00}. The main difference with OSS seems to be ability to a) leverage an existing community of enthusiasts, and b) gain full access to the codebase.

Forking is relatively uncommon in open-source projects, perceptions notwithstanding. Two primary reasons are the open, accessible nature of OSS licences, and the social cost in splitting from an existing project. An example illustrates the first reason \cite{moen99}. The {\tt gcc} compiler team had decided not to support particular optimizations; another team forked the codebase, then went ahead and implemented the optimizations.  This alternative proved popular, and the resulting project, {\tt egcs}, was eventually merged back into the {\tt gcc} tree.  The second reason, social support, detailed in Raymond (\cite{raymond01}, p. 84-87), is more common.  Once a community has developed around a tool, it is difficult to create a variant of that tool and maintain the critical mass needed to keep the quality high.  These concerns also appear in the qualitative studies we conducted.
    
\subsection{Requirements evolution}
\label{re}

Another important context for this research is the evolution of requirements. Understanding one's requirements is central to understanding what features a software system should implement.  Where those requirements overlap with existing projects, the opportunity for a fork exists. Forking software to meet new requirements is equivalent to evolving an existing software product.

For {\tt DrProject}, there was no initial explicit requirements analysis.  Rather, as one of the {\tt DrProject} proponents mentioned, a ``chasing tail-lights'' approach was used.  This means identifying a product that does something similar, then molding your tool after that.  In this sense, many OSS projects can be seen as `forks' from existing tools, in ideas if not actual code.  

There are three possible avenues to satisfy the identified requirements (regardless of how they are identified or specified).  One, the proponents can create an entirely new project from scratch. This might include the use of commercial off-the-shelf products. This was tried, unsuccessfully, with earlier ancestors of \dpr. Two, the proponents can identify existing tools that partially satisfy the requirements, and work within that community to fill the unsatisfied requirements. To be successful such an approach involves convincing a community of the merits of your needs. In the case of \dpr, the proponents felt this was unlikely, given the different goals they had. Finally, the proponents in OSS projects can fork the code and adapt it on their own to create a variant that meets their needs. This last choice was made for \dpr. Georgas, Gorlick and Taylor \cite[p. 3]{georgas05} describe why this last choice makes sense: ``In the world of open-source software development, the chances are excellent that someone else, somewhere else has already solved your problem, in which case, it is a foolish waste of effort to solve it again''.

\section{Research design}
The initial review of the literature produced little in the way of clear theories about forking and open-source software. To address this, we chose a mixed-methods research framework for the project. By way of introduction to this approach, we highlight some of the discussion regarding the methodology from Creswell's book on research design~\cite{creswell02}. 

Mixed-methods approaches combine both qualitative and quantitative research methods.  This combination can occur either in parallel or in sequence. Primarily this is done to combine the best of both approaches: the open-ended, generative nature of qualitative research and the confirmatory nature of quantitative studies. Mixed-methods research techniques arose from research in psychology, and an interest in mixing different data sets. It is growing in popularity in sciences which straddle physical/social boundaries, such as psychology, economics, geography, and, we would argue, requirements engineering, with its strong human-centered focus. Mixed-methods research is a methodology with a strong pragmatic motivation. This can be contrasted with the constructivist approach common to many qualitative studies, and the post-positivist, rationalist approach common to quantitative studies.  Pragmatism focuses on real-world implications and consequences of action~\cite{creswell02}.

\begin{figure*}[htb]
\begin{center}
\includegraphics[bb=0 0 769 277,width=0.8\textwidth]{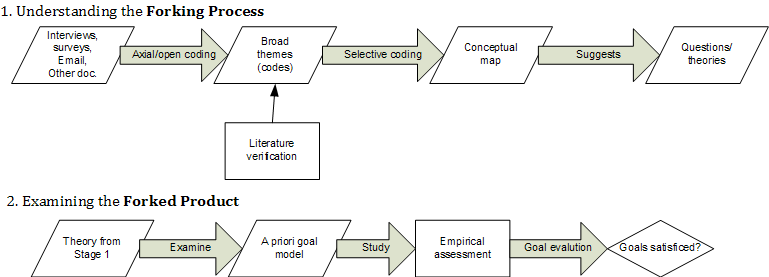}
\caption{Process model of the two phases of the research design. Phase 1 is a qualitative theory generating step about the process of forking. Phase 2 takes one theory about this process and applies it to a product of the process.}
\label{fig:method}
\end{center}
\end{figure*}

The challenges of mixed-methods research include the need to understand both quantitative and qualitative procedures, and the longer time frames possible (particularly in sequential studies, where data from one inquiry, such as the qualitative work here, inform the next stage).  However, we have found that using the qualitative inquiry successfully narrowed and informed both the choice of quantitative method and the questions to focus on. It works well when there is little existing research to form well-grounded theories about a problem. Figure \ref{fig:method} illustrates the combined approach we took. 
\section{First stage: qualitative inquiry}
This phase was a qualitative inquiry into the nature of the process of forking in open-source software. The results of this inquiry are questions, stated as theories, about this process.

\subsection{Methodology}
\label{method}
We gathered data from both sets of groups involved in the fork. We used a survey\footnote{Questions can be found at \url{http://neilernst.net/docs/trac-survey.txt}.} to gather responses from the developers most involved with the {\tt Trac} project. The survey was emailed to five individuals; by the deadline, three responded.  All three were senior developers, although not the `gatekeepers', for the {\tt Trac} project. For the \dpr\ team, we interviewed both leaders in person, transcribing the interview from the audio file. We also examined mailing list records for each project.

Having gathered this dataset, we then used grounded theory methods of open, axial, and selective coding to generate concepts (theories) regarding the problem.  `Coding' a set of data involves establishing commonalities that emerge (open coding); grouping these commonalities, or concepts, into higher-order clusters (axial coding); and finally, establishing potential relationships among the concepts (selective coding) \cite{strauss98}. To do this, we read through the various source materials (emails and transcripts), highlighting terms and ideas that seemed salient.  Those which appeared in a majority of sources we used as codes. These codes are discussed at length in Section \ref{qual}. Note, we used all the interviews to generate the codes.  Our aim was to generate codes which were grounded in both sets of experiences. Codes in this context can be seen as dimensions, or themes, regarding the process of forking and open-source software.

\begin{figure*}[htb]
\begin{center}
\includegraphics[bb=0 0 769 438,width=\textwidth,height=3in]{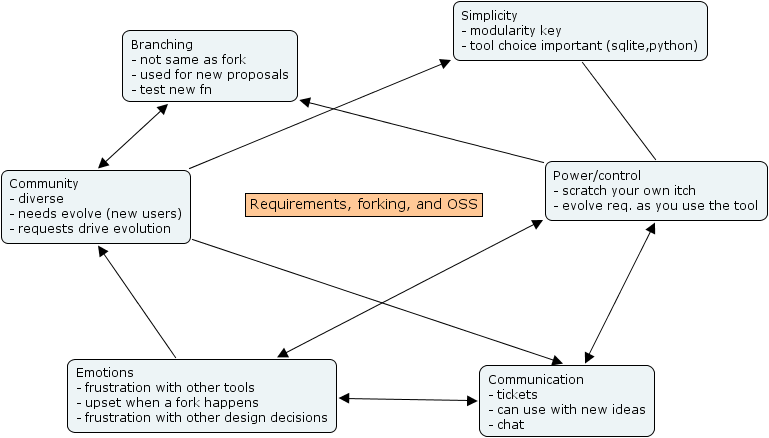}
\caption{Relationships and codes used in theory construction}
\label{fig:code}
\end{center}
\end{figure*}

\subsection{Observations}
\label{qual}
Figure \ref{fig:code} illustrates the six codes/concepts that we extracted from the data. Again, these represent common themes presented in the data. After deriving them, we returned to the literature to confirm, as best possible, that the dimensions we identified were not contradicted by the literature. Below, we present the code we generated, along with summaries of the type of data that gave rise to the code.
\begin{description} 
  \item [Branching] -- using branches to promote alternate visions for a tool; how a fork differs from a branch.
  \item [Community] -- discussion on diversity of the community; how community requests are driving requirements; how the tool evolves as more people adopt it.
  \item [Simplicity] -- the idea that code should be modular; that student projects require less complexity due to shorter time frames; that chosen technology ({\em e.g.}, language choice), if simple, can help attract other developers and users.
  \item [Power/control] -- the `scratch your own itch' meme; that requirements evolve as the tools are used; that different requirements can imply different directions; that a good community has a `gatekeeper' to oversee development and drive the project.
  \item [Communication] -- tickets as a source for new ideas and a way of keeping the community informed; concern over appearing closed off, using IRC or chat, which have no easily accessible logs.
  \item [Frustration/emotions] -- passion over existing tools as inadequate; emotion over a fork resulting in your hard work supporting something you have no power over; frustration over design decisions in another project; unease over project directions.
\end{description}
As the figure indicates, these can be connected selectively to some of the other concepts.  \textbf{Branching} helps with community building; other people can get involved without having to convince others of the benefits of their project until they are ready to merge their changes.  This is related to a power/control concept as well.  \textbf{Simplicity} helps to build a community, allowing others to develop software more easily. This allows for people to take control of the issues they are interested in. For example, {\tt DrProject} is designed to be simpler than \tra\ so new developers can quickly understand the project. This is because \dpr\ is intended for student projects and student maintainers. \textbf{Community} needs, while facilitated by simplicity and branching, can involve emotions, including frustration, and requires good communication support. \textbf{Power and control} issues can be supported by effective, early communication about what requirements should be addressed; this issue also can lead to frustration, for example if a requirement is not considered important, or code one person worked on is forked to a different project where that individual has no influence.  This is perhaps the most complex of the issues generated by the survey, and requires more investigation.  \textbf{Frustration} is a product of failing to address the other concepts.

At this stage, we have taken the various data elements, generated common concepts, and then created relations among these codes selectively (Fig. \ref{fig:method} illustrates this process). Once we have these dimensions in the form of the concept map in Fig. \ref{fig:code}, questions can be posed regarding the process of forking and open-source software. We went through the concept map, deriving questions suggested by the codes. We present these questions below in the form of simple theories regarding the process. This list is not exhaustive, but does capture many salient aspects of the process. We present a keyword to identify the theory, followed by its description. In brackets, we identify the relevant codes which give rise to that theory.
\begin{description}
\item[Divergence] -- A forked project will diverge more from the main codebase over time than a branched subproject will [Branching, Power/control].
\item[Formality] -- OSS projects typically consider requirements informally, using branches and feedback to suggest new features [Community, Branching].
\item[Activity] -- In community-developed projects ({\em i.e.}. OSS), there will be a high number of branches with frequent commits, representing development on new requirements/features [Community, Branching, Power/control].
\item[Definition] -- Projects with well-defined requirements will have fewer developers and fewer branches, and less complex code [Simplicity, Community, Branching].
\item[Leadership] -- Central to every project is the requirement for a leader who has a good grasp of the surrounding issues of control, community, and communication [Power/control, Frustration].
\item[Openness] -- Publicly accessible chat logs will lead to more open communication about a project, and produce less dysfunction in community building [Simplicity, Community].
\item[Modularity] -- The more modular a tool is, and the more it uses well-known tools, the more likely it is to be adopted [Simplicity, Power/control].
\end{description}
We discuss some ways to explore these theories in more detail in Section 5. Given scope constraints, we chose one additional theory to develop in more detail. This was generated in the same way the others were, but we focus on it here to emphasize its use in the following section (as per Fig. \ref{fig:method}. This theory questions why the decision to fork might be made.
\begin{description}
\item[Motivation] -- Forking occurs when the requirements of a system-to-be extend beyond those of an existing system, and there is an inability to address those new requirements within the existing project [Power/control, Community, Simplicity].
\end{description}
In the specific case of \tra\ and \dpr, the \dpr\ team understood priorities for \tra\ to be different. For example, the \tra\ team were unable to commit to supporting the integration of multiple-project support. Bonaccorsi and Rossi\cite{bonaccorsi05} describe a similar occurrence in the context of corporate participants in OSS projects. They tended to extend existing projects to address customer needs. These two examples lend some rationale support to this theory. Because \tra, for one, was fairly well-structured, it was relatively simple to adapt the existing tool to new requirements. Most of these requirements arise from the new context -- supporting undergrad projects within term-length courses, administered by course instructors and students. 

\section{Second stage: confirmatory analysis}
\label{quant}
This second phase took the hypothesis generated in the preceding stage and conducted empirical tests. Primarily these consisted of quantitative testing; however, we also used qualitative user testing in the second portion, since this test applies well to the chosen hypothesis. 

The original theory ({\bf Motivation}) centered on the question of deciding to fork. We suggested that this is done when one group identifies additional requirements that the existing tool cannot, or will not, address. To test this theory, we want to identify several factors. 
\begin{itemize}
\item One, were there additional requirements, and were they not being met in the original context? That there were additional requirements is shown by the goal model of Fig. \ref{fig:goal}. That these were not addressed by the \tra\ project is argued in the preceding section. 
\item Two, were these additional requirements addressed in the new, forked project? If not, then the rationale for forking seems less clear.
\end{itemize}

In this section, we tackle this question of whether forking \tra\ was successful in meeting the new requirements. Recall the goal model of Fig. \ref{fig:goal}. In goal modeling terminology, we wish to assess whether the new requirements are satisficed by the new system. In particular, we focus on two non-functional goals (softgoals), namely usability and maintainability. These were identified {\em a priori} as important characteristics for the system-to-be to satisfice.

\subsection{Methodology}
To assess whether the forking achieved its objective, we can measure success for two of the most important requirements: long term maintainability by instructors/students (using code complexity as a key indicator), and minimal learning time for undergrads (using usability testing on paper prototypes as an indicator). We do not need to provide proof this was done. Indeed, since these two requirements are non-functional, by definition we can never know whether they are fully satisfied. Instead, this study seeks to provide sufficient rationale to confirm both goals are satisficed by the new system (satisficing indicates that we can be reasonably content with the outcome \cite{simon67}).

Both requirements can be analyzed empirically. First, we derive a hypothesis to guide this phase of the research. It is:

{\bf H$_1$} -- {\em The softgoals of maintainability and usability that partially motivated the decision to fork {\tt DrProject} have been satisficed}. 

We formulated several techniques for testing the hypothesis.  Below, we outline each approach and the associated claim to be investigated. Results are presented in the following section. 

\subsubsection{Maintainability}
\label{mccabe}
We used the {\tt pymetrics} tool\footnote{http://pymetrics.org} to assess software maintainability. We normalized the two projects by removing any testing code as well as those python files with fewer than fifteen lines of code (such files consisted of initialization code not central to program comprehension).  This left all python source files that contributed directly to the functionality of each project.

\begin{enumerate}
  \item We generated McCabe cyclomatic complexity~\cite{mccabe89} figures for both the {\tt Trac} and {\tt DrProject} codebases. This metric calculates the number of paths through a program call graph, measuring decision points ({\em e.g.}, if statements) and exits. It is used to assess the relative complexity of a codebase. For example, more decision paths branching from a particular point in the code indicates that maintenance of that portion of the code may be troublesome (as changes have greater impact). 

McCabe may not be the best metric for dynamically typed, web-based systems like \dpr. Muthanna {\em et al.} \cite{muthanna00} suggest it correlates well with maintainability, but they studied larger C language systems. Rajaraman and Lyu~\cite{rajaraman92} conclude the McCabe metric doesn't work well for object-oriented C++ systems, but \dpr\ and \tra\ exhibit few strong OO tendencies (such as encapsulation or inheritance). We conclude that while imperfect, the McCabe metric can provide evidence towards concluding whether or not goal satisficing is achieved.  

{\bf H$_1a$}: \tra\ code is less maintainable than \dpr\ according to the McCabe cyclomatic complexity metric.  

	\item We performed a lines of code (LOC) analysis.  This is a weak metric for software complexity, but provides another data point. Its most obvious weakness is that what one programmer does in ten lines, another may do in two (and neither number says much about software quality). It nevertheless provides a quantitative assessment of the absolute size of the codebase, and in general, larger codebases are more complex. One aspect of the codebases we haven't considered is output handling, that is, the code which controls output to the browser. We discuss this qualitatively in the following section, under {\em type conversion}.

	{\bf H$_1b$}: {\tt DrProject} has fewer lines of code than {\tt Trac}.

	\item We assessed the extent of commenting in each project. A key factor in complexity and maintainability is the degree to which source comments are used. Poorly commented code is more difficult to maintain without assistance from the original author. In Python, comments can be added at the function and class level using `doc-strings', which are similar to Java's `Javadoc' mechanism. We measured the proportion of functions with comments.

	{\bf H$_1c$}: {\tt DrProject} has a greater proportion of functions with comments than {\tt Trac}.
\end{enumerate}

\subsubsection{Usability}
As a final means to evaluate {\bf H$_1$}, we introduce the results of qualitative usability testing of a revised {\tt DrProject} interface. The new interface was a redesign of the existing \tra\ interface to address workflow issues introduced by several new modules. We used paper prototypes \cite{snyder03} of proposed interfaces to assess usability. We interviewed five participants, assigning them a series of tasks to perform in {\tt DrProject}. For example, one task was to log in to the system. After completing the tasks we presented them with a series of questions, in a structured interview format. We used this to redesign the interface and tested this updated version with a further five students.

\subsection{Observations}
\subsubsection{Maintainability}
\emph{Code complexity} -- We ran the pymetrics tool on the source files, producing individual results for each one.  We then averaged these results to obtain a project McCabe complexity metric. We also computed the COCOMO2 SLOC metric \cite{boehm00}, which measures single lines of code (excluding comments and empty lines). This we aggregated by summing all the values to produce a total value. Finally, we generated a measurement of the proportion of functions which were commented in the source. These were also averaged to produce the mean value. Results are shown in Table \ref{tbl:cc}. 

{\tt DrProject} is 27\% less complex according to the McCabe metric.  It also has over twice as many functions with comments, enhancing maintainability. Lines of code are marginally lower in {\tt DrProject}. However, this number includes two additional features not currently part of the production system. These are a requirements and testing component and a graphical dashboard. These add over a thousand lines to the {\tt DrProject} codebase.

 \begin{table}
\begin{center}  % put inside center environment
\begin{tabular}{|c|c|c|}
\hline % note: updated for DP2
  \textbf{Metric} & \textbf{\tra} & \textbf{{\tt DrProject}}  \\
  \hline \hline % put a line under headers
  McCabe ($\mu$) & 56.25 & 44.49  \\
  \hline
  SLOC (sum) & 7572 & 7504\\
  \hline
  F$^n$ Comments & 21.6\% & 46.2\% \\ \hline
\end{tabular}
\caption{Aggregate normalized software metrics for the two projects}
\label{tbl:cc} 
\end{center}
\end{table}

\emph{Type conversion} -- Type conversions increase cognitive load when data is translated from one form to another. Programmers must remember both the new and existing form, as well as a new reference to the data. We examined the frameworks chosen by both \dpr\ and \tra\ to assess type conversions. This was prompted by a claim by one of the leaders of \dpr\ that the number of type conversions was lower in the new codebase. A framework is a tool, or suite of tools, that make it easier to accomplish certain tasks. In this case, the principle framework difference is in HTML display. \tra\ uses the Clearsilver\footnote{http://www.clearsilver.net} tool for producing output. Its advantages are speed and cross-language compatibility. The \dpr\ team moved to Kid\footnote{http://lesscode.org/kid}, which is only for the Python language, and is designed for producing valid XML (of which XHTML is one language). Inspecting the differences in these two languages shows that Kid reduces type conversions, since it is a native Python tool. Data remains in the same form in either a Kid template (the view) or Python source file (the model). Clearsilver, by contrast, uses a hierarchical data model that requires the user to insert data for display by converting to a string, then extract it for use in the template. Another benefit is that Kid is explicitly designed for XML production, and is easier to secure against injection attacks. Finally, since Kid insists on valid XML, there is no possibility of generating invalid XHTML. These changes contribute positively to the satisficing of the maintainability NFR.

Given the results of the preceding three sections, we conclude that hypothesis {\bf H$_1$} provides a sufficient explanation of the data; namely, that the maintainability softgoal is satisficed.

\subsubsection{Usability}
While we cannot conclude \tra\ is less usable than \dpr, the iterative redesign does account for the different requirements of the \dpr\ tool, namely that a student user group needs a simpler interface. For example, \dpr\ has added multiple project support and substantially reduced interface complexity. The ticket interface, for example, has reduced the available options by three fields (from nine), and the available values for those fields by nine (from eighteen). However, there remains a tension between the functional features in the UI -- such as modal interface to support multiple projects -- and the softgoal of usability. While our subjects had little problem completing our walkthrough, they ran into difficulty with how the various projects interacted. For example, one user commented that he wanted ``tickets to show up across projects''. Another user, however, liked the custom ticket interface we presented her.  \footnote{Data available on request}. The usability testing needs to be more extensive to make any firm conclusions. This is a well-evaluated area of HCI. However, it would be difficult to do relative comparison of two interfaces over the time period required for this study. Ethics concerns would prevent the use of two tools of (hypothetically) differing quality. Students use \dpr\ for projects, and we wouldn't want to impact marking. However, the qualitative assessment we undertook does provides some support for hypothesis {\bf H$_1$}.  We conclude that the usability softgoal is partially satisficed.

\section{Discussion}
\label{implications}
We begin with an analysis of the chosen research methodology, including threats to the research validity.  Then, we analyse the implications of the study results, including some notes on additional theories and ways to test them.

\subsection{Research validity}
\label{threats}
Good empirical studies include an assessment of how applicable the results are, and how well-founded any conclusions may be. The following discussion is modeled after suggestions in Trochim~\cite{trochim01}.

\emph{Conclusion validity} --  Conclusion validity assesses whether the conclusion we drew was supported by the evidence. We concluded that there was a relationship between differing requirements and {\tt DrProject}'s usability and maintainability. The largest threat here is the use of code complexity metrics, as discussed in Section \ref{mccabe}. These provide a concrete number to which it is easy to assign more meaning than appropriate. Similarly, there are no doubt more appropriate metrics -- in general, there is a scarcity of suitable metrics for web application frameworks such as those used here. Such metrics would assess factors like framework complexity (J2EE vs Rails), type conversions, etc. Nonetheless, we maintain that the metrics generated do point to substantive differences in maintainability. While not definitive, they do allow us to claim that the maintainability softgoal is satisficed. The qualitative procedures, while adequate, could have been broadened; the sample sizes were somewhat small (although, we would argue, representative for this project). For more general results, better theoretical sampling is required. 

There is a question as to whether there is any possible failure associated with this hypothesis (that is, whether it is tautologous). One might argue that having identified the requirement of, for example, reduced complexity, an outcome of reduced complexity is nothing surprising. We argue differently. During the timeframe of this study, there were other development goals beside the two softgoals we list. For one, all the functional requirements, such as multiple-project support, needed to be addressed. Second, there were students involved in the project who had competing requirements. These were students funded to develop research prototypes, such as project dashboards, that were not part of the core \dpr platform. These tools distracted from the core purpose of the development effort. Accordingly, there was a very real possibility that the new tool would be neither more usable nor more maintainable.

\emph{Internal validity} -- Was the relationship established a valid one? There may have been hidden factors influencing the results, but properly conducted grounded theory research~\cite{strauss98} demands that the researcher examine his or her own sources of bias during the research, in order to accommodate this. This we did by introspection of the methodology and results.

\emph{Construct validity} -- This measures the extent to which the theoretical constructs applied were valid. In this case, the study was largely exploratory; the correct techniques and definitions are as yet largely undefined. We defined two construct, usability and maintainability. The measures were quantitative assessments of the relative degrees of complexity of each. There was some degree of hypothesis guessing on behalf of the survey participants, who attempted to guide conclusions. We addressed this by using only codes which emerged from a majority of the data sources, not just one or two. Finally, codes should be generated by more people to compare the inter-rater reliability of the results.

\emph{External validity} -- Qualitative studies can be hard to generalize. However, we believe the results suggest several interesting research opportunities, as the \tra\ project seems characteristic of many smaller open-source projects. How well this maps to larger projects, for example Apache, is uncertain. The only generalization we make is that the ability to fork a project to meet requirements which are a superset of the existing ones seems possible given the experiences reported here.  However, since this is but one of many such examples, more research is needed before concluding this is the case.

\subsection{Results of the fork}
The confirmatory testing conducted allows us to conclude that {\tt DrProject} has partially satisficed the usability softgoal, and satisficed the maintainability softgoal. Despite adding several significant functional enhancements, the codebase is less complex than the existing \tra\ code (when one might expect it to be as complex or more complex). It has a student user-tested interface. The code is better documented. Finally, although not explicitly verified, the code has more testing infrastructure to verify software correctness. The fork has been successful in adapting existing features to new requirements. According to our initial, grounded theory of {\bf Motivation}, which questions what motivates a fork, we can conclude that a) there was good rationale for forking; b) the forked product addressed new requirements successfully. This implies that, in this case, forking was an acceptable means of addressing the requirements identified previously.

What were the reasons for this? We identify five factors.
\begin{itemize}
  \item There was a good match between the core requirements ({\em e.g.}, repository integration, bug tracking, etc.). Indeed, the {\tt DrProject} proponents modeled many of the requirements after \tra\ or similar tools.
  \item Existing code, while poorly documented, was well-structured and architected, and
  \item One of the {\tt Trac} architects was willing to help transition the codebase during the fork period, answering questions and providing technical assistance for a period;
  \item The {\tt DrProject} initiative had strong leadership with a clear vision of what the tool should be, and dedicated developers during the initial period of the fork;
  \item Finally, lessons learned over the past years provided valuable experience on what a manageable set of functionality was.
\end{itemize}
Does this imply that the technique of forking is a viable one for other domains? One could argue that having a strong leader and clearly defined requirements meant that the {\tt DrProject} effort would succeed regardless.  There are two counter-arguments.  One, the same leadership had attempted virtually the same project several times in the past, with little success. A strong vision alone is not sufficient. Secondly, the scope of the project (as defined by the perceived requirements) was such that the four month development period would have been insufficient to fulfill all those requirements. 

We suggest that what occurred was that this process was actually the evolution of an existing system into a modified system, reflecting new external forces (requirements changes). The implicit requirements model was revised accordingly. Forking the code therefore leverages the need for a first, `throwaway' implementation \cite{brooks95}. Assume the requirements for two systems are relatively similar. The newer system can build upon those other requirements ({\em e.g.}, using pre-existing repository integration to build advanced visual interfaces). This reduces the amount of time required to code an entire system.  Forking becomes a software (and requirements) evolution and maintenance task. The disadvantage is that earlier developers typically will not assist in the process, but this is often the case in software evolution, where previous developers have moved on.

\subsubsection{Costs and benefits of the fork}
Concluding that \dpr\ met the requirements of improved maintainability and usability for the fork suggests forking was beneficial.  However, it is possible some costs were incurred as well. To investigate what these might be, we conducted a second set of interviews with the {\tt DrProject} leaders. The interviews were structured along the five concepts described in Section \ref{qual}. We used these concepts to guide a separate discussion with each of the two leaders. From transcriptions of these interviews, we elicited the following `concerns' about forking software.
\begin{description} 
  \item [Branching vs. forking] -- The fork moved code out of the {\tt Trac} repository, but this wasn't felt to be a major cost. Even had the project remained in the repository (branched), the differences in requirements would have made it very difficult to automatically merge code into the mainline. There was some concern that change the other way would be difficult, e.g., getting the latest bugfixes and security patches from {\tt Trac}. 
  \item [Community] -- Forking has meant the loss of the {\tt Trac} community of developers. There are plans, however, to allow other schools to use the tool and possibly contribute to its development. In the short term, though, there are fewer users to help identify bugs and fix problems. The tool seems out of reach for undergraduates to participate in -- certainly for a senior programming assignment. A student might be happy to receive 80\% on an assignment, but to one {\tt DrProject} leader, ``I regard that as 20\% of your code is buggy \dots which is just too high''. 
  \item [Power/control] -- Forking the codebase has meant the two {\tt DrProject} proponents have had to undertake the leadership of the new project, with concomitant responsibilities for management and future development. There is a hope that other universities will join in, but some involvement from these two seems likely in the immediate term. Tool documentation and code stability are not a point where sharing the tool with others is feasible. Finally, some issues will remain local, e.g., authentication processes are controlled by the organization's existing processes.  
  \item [Communication] -- There was some cost to the community associated with a less open communications leading to less visible decision-making. Logging of IRC meetings is done, but it would be ``wrong-headed ... to shift communication'' to something the core team wasn't comfortable with. As with all OSS, however, the fact the code is accessible means a {\em de facto} level of openness that non-OSS projects don't exhibit.
  \item [Frustration/emotions] -- There has been no evidence of negative feedback from the existing {\tt Trac} community. There are two suggestions for this. One is the work by one individual on both {\tt Trac} and \dpr\. This person is a key contributor, and his involvement lends credibility to a project he is involved with. In this case, this implies a validation of the differing requirements for the new project. The second factor is that the reason for the fork was primarily for technical reasons, and not personal ones. This seems to be a less emotional process, where disagreement, if it is found, is principally on technical grounds.
\end{description}
Of these, the factors identified as most costly were the loss of bug fixes from the community, and the potential gap in creating a new community.

\subsection{Assessment of the research technique}
This study used a sequential `explore and confirm' mixed methodology to explain and understand the nature of forking open-source software. This technique was useful in generating theories about the problem where the literature didn't provide much insight.  Many of these theories, however, defy quantitative evaluation.  For example, how can we understand how a particular project considered requirements other than qualitatively? We avoided this dilemma by focusing on the quantifiable aspects of the system, such as code metrics. There remain many interesting issues, such as team dynamics, collaborative work, and communication, that are best assessed qualitatively.

Another approach to evaluating forking (and system evolution in general) would consider other non-functional requirements \cite{mylopoulos92} such as testability, modularity, and maintainability. Techniques similar to those mentioned in Yu {\em et al.} \cite{yu05b} could be applied: reverse engineer a goal model from the existing source; elicit new requirements for the potential fork; apply tradeoff analysis to derive a new goal model. 
 
\section{Conclusions}
\label{conclusions}
This research explored theories about forking, requirements, and open-source software. There were two problems addressed by the work. Firstly, what is the nature of the forking process? We identified some questions in this area such as the nature of requirements in forking, the attitude between the communities, and whether requirements were a basis for a fork. Secondly, we identified one question, regarding the motivation to fork. We hypothesized that forking was required to address the softgoals of maintainability and usability. To confirm this, we showed that the new codebase, while a better match for the new functional requirements, satisfices the usability and maintainability softgoals.  Given the relative success of the DrProject fork -- adapting an existing tool to new requirements -- we conclude that this is a potentially useful avenue in software development. We then analysed some threats to the validity of the research, and discussed the findings and their implications.  The theories generated in the first stage suggest several possible areas for future research in this problem domain. 

\section{Acknowledgements}
We would like to thank Greg Wilson, Karen Reid, and the students who worked on {\tt DrProject}, as well as the Trac developers who took the time to respond to our survey. Finally, thanks to Jen Horkoff and Jorge Aranda for their insightful reviews.

\section{Biographies}
{\em Neil A. Ernst} is a Ph.D. candidate in software engineering at the University of Toronto. His research interests include requirements engineering, information visualization, and human factors.

{\em Steve Easterbrook} is Professor of Computer Science at the University of Toronto. His current research goals focus on the analysis of requirements for complex software-intensive systems.

{\em John Mylopoulos} is Professor of Computer Science at the University of Toronto. His research interests include information modelling techniques, covering notations, implementation techniques and applications, knowledge based systems, semantic data models, information system design and requirements engineering.

\bibliographystyle{IEEETran}
\bibliography{neilernst}

\end{document}